\documentstyle[preprint,pre,aps]{revtex}

\begin{document}

\draft

\preprint{Version of 6/27/94}

\title{
The Inhibition of Mixing in Chaotic Quantum Dynamics}

\author{B. S. Helmkamp and D. A. Browne}

\address{
Department of Physics and Astronomy\\ Louisiana State University\\ 
Baton Rouge LA 70803--4001}

\date{}

\maketitle

\begin{abstract}
We study the quantum chaotic dynamics of an initially well-localized
wave packet in a cosine potential perturbed by an external
time-dependent force.  For our choice of initial condition and with
$\hbar$ small but finite, we find that the wave packet behaves
classically (meaning that the quantum behavior is indistinguishable
from that of the analogous classical system) as long as the motion is
confined to the interior of the remnant separatrix of the cosine
potential.  Once the classical motion becomes unbounded, however, we
find that quantum interference effects dominate.  This interference
leads to a long-lived accumulation of quantum amplitude on top of the
cosine barrier.  This pinning of the amplitude on the barrier is a
dynamic mechanism for the quantum inhibition of classical mixing.

\end{abstract}


\narrowtext

\section{Introduction}
\label{sec:intro}

While the inhibition of classical mixing in quantum chaotic dynamics is
a generally accepted consequence of the linearity of the Schrodinger
equation, how this inhibition is manifested in the dynamics of physical
systems is not fully understood.  This is a difficult question to
address from a theoretical point of view because the complex nature of
chaotic systems all but precludes an analytical approach, making
general physical insights difficult to come by.  However, by studying
the semiclassical limit for simple low-dimensional systems, one can
relate the quantum dynamics for specific case studies to generic
features of the classical chaos and thereby discern generic
semiclassical mechanisms that lead to the quantum inhibition of
mixing.

The mechanism responsible for classical mixing is the repeated folding
and stretching of the classical Lagrangian manifold in the vicinity of
hyperbolic fixed points \cite{Berry}, and when the motion is chaotic,
these folded structures, or ``tendrils,'' eventually fill almost all
the available phase space.  The folding and stretching process leads to
an abundance of caustics, or turning points, in the Lagrangian
manifold. Since quantum effects are, loosely speaking, amplified at
caustics, this can have a dramatic effect on the semiclassical
description of the motion. One example of this is seen in the
scattering of a wave packet off of a barrier\cite{Schiff}.  When the
wave packet reaches the turning point, it acquires a temporary standing
wave modulation arising from the interference between the piece of the
packet that has already reflected off the caustic and the piece that
has not yet reached it.  Thus, the caustic temporarily amplifies the
underlying quantum nature of an apparently classical object.  Schulman
quantified this propensity for nonclassical effects at a caustic by
showing that the contribution to the propagator per degree of freedom
should go as $\hbar^{1/2}$ rather than $\hbar ^{1/3}$ \cite{Schulman}
within a ``critical region''\cite{areahbar} of the caustic where the
actions of the direct and reflected paths differ by less than $\hbar$.

Despite the dynamical sense conveyed by the term ``mixing,'' most
studies of the inhibition of mixing in quantum chaos have explored
either the statistics of energy levels or the spatial localization of
eigenstates for (bounded) conservative systems.  Localization, or
``scarring,'' of chaotic eigenstates was first observed by Heller for
the stadium billiard \cite{stadium}, a strongly chaotic system with
isolated unstable periodic orbits. Heller's scars are local
enhancements above the background intensity that occur along those
orbits that are the least unstable---meaning that the ratio of the
frequency of the orbit to the Lyapunov exponent that characterizes the
divergence of adjacent trajectories is significantly greater than one.
A second mechanism for eigenstate localization has been observed by
Bohigas {\em et al.}\cite{Bohig} in the soft chaos of two weakly
coupled quartic oscillators. The local enhancements found in the
eigenstates for this system arise from the quantization of the remnants
of Kolmogorov-Arnol'd-Moser (KAM) tori, or ``cantori.''

In contrast, we focus on the time domain rather than the energy domain,
using semiclassical ideas to relate the quantum dynamics to the
evolution of the classical Lagrangian manifold for a chaotic
Hamiltonian system.  In a recent publication\cite{me} we discussed the
role of classical phase space structures in the quantum evolution of a
wave packet prior to the onset of chaos.  In particular, we considered
the motion of a quantum particle in a cosine potential perturbed by an
external force varying sinusoidally in time. We posed the problem in
the naive semiclassical limit where the natural length, mass and time
scales make $\hbar$ small.  We studied the dynamics of a wave packet
that was initially well localized in position and momentum and compared
it to that of the classical distribution in phase space. Provided the
classical distribution remains inside the separatrix of the unperturbed
motion and avoids the hyperbolic fixed points on the separatrix, the
two initially identical distributions were found to remain
indistinguishable, even though both spread out to fill much of the
initial well. However, when the Lagrangian manifold interacts with a
hyperbolic fixed point and acquires a tendril-like feature, the wave
function was found to exhibit a persistent nodal structure that
represents the first appreciable quantum effect with no classical
analog.  We also showed that this feature is relatively benign, having
little impact on the physical observables.

In this paper we consider essentially the same system but focus our
attention on the classical exit event when a substantial fraction of
the classical distribution crosses the separatrix and leaves the well,
marking the onset of chaos. Here we will show that the apparent
equivalence of the quantum and classical distributions rapidly
disintegrates as the caustic in the Lagrangian manifold moves past a
hyperbolic fixed point, {\em i.e.}, as the turning point moves outside
the initial well.  Specifically, we observe that while the peak in the
classical probability density associated with the caustic follows the
caustic out of the well, the peak in the quantum distribution becomes
pinned at the top of the barrier.  This pinning effect represents a
specific dynamical mechanism for the quantum inhibition of classical
mixing in a time-dependent Hamiltonian system.

We have deliberately chosen to study a time-dependent Hamiltonian with
one degree of freedom because it is sufficiently complex to exhibit
chaos yet simple enough to be studied directly in phase space.  We
point out that this problem is not equivalent to the conservative
two-dimensional problem obtained by elevating time to the role of an
additional degree of freedom because of the distinct role of time as a
parameter rather than an operator in quantum mechanics.  The
quasi-conservative system that one obtains by invoking the strobed-time
Floquet formalism\cite{floquet} is not relevant to our problem either
since the phenomena we observe occur on a time scale that is a fraction
of the period of the driving and cannot be discussed meaningfully in
terms of the quasi-energy Floquet states.  Not being constrained by the
symmetry of a time-invariant Hamiltonian, the mechanisms we observe are
not limited to conservative systems and may, in fact, be disallowed by
that symmetry in certain cases.

Although the origin of the pinning effect, like the nodes of our
previous paper, can be understood semiclassically, the origins of the
two effects are distinctly different.  The nodes in
Ref.~\onlinecite{me} were shown to be the result of a beating
phenomenon in the Van Vleck-Gutzwiller (VVG) propagator\cite{vvg}
between paths having the same Gutzwiller phase.  The most prominent
nodes were found to be the closest in action to a false caustic in the
flow field which developed as result of the interaction of the
Lagrangian manifold with the hyperbolic fixed point.  In contrast, the
structure we observe here is a result of interference between direct
and reflected paths---which differ by $\pi/2$ in Gutzwiller phase---and
the most prominent structure is in the neighborhood of a true caustic.

Because the stationary paths in the path integral converge at a
caustic, the stationary phase approximation that leads to the VVG
propagator by summing each classical path independently breaks down.
Therefore, to discuss the behavior near the caustic, we adopt a
semiclassical propagator derived by Schulman\cite{Schulman} that is
strictly valid only in the immediate vicinity of a caustic.  Since this
propagator cannot be evaluated numerically as easily as the VVG
propagator, we develop an approximate ``connection formula''--- similar
in spirit to the WKB connection formulae---to evaluate the Schulman
propagator from the VVG expression.  This yields an expression that is
valid even at the caustic.  We use this hybrid propagator to show that
the differences between the classical and quantum behavior associated
with the classical exit event can be understood semiclassically in
terms of the area-preserving deformation of the manifold.

Furthermore, our propagator can be used to study the exponential tail
of the wave function in the shadow or classically forbidden region of
the caustic which is missed completely by the VVG propagator. This is
important to our problem since the stretching of the exponential tail
is the primary semiclassical mechanism for the quantum system to
explore the world outside the remnant separatrix. Hence, the VVG
propagator, which puts zero amplitude in the classically forbidden
region, becomes inadequate to describe the dynamics of the wave packet
after the exit event. We point out that this failure of the VVG
propagator is not necessarily inconsistent with the long-time accuracy
achieved by Heller and Tomsovic \cite{unexp} in the case of the stadium
billiard. Since the amplitude in the forbidden regions is evidently
negligible in that problem \cite{areahbar}, it follows that the
stretching of the exponential tail and associated phenomena that we
observe must not be significant there.

The main body of this paper is organized as follows. In Section
\ref{sec:problem} we describe the Hamiltonian, the choice of initial
conditions and the method of analysis.  (The present discussion is
rather brief, since a complete discussion is contained in Secs.~II and
III of Ref.~\onlinecite{me}.) We explain our semiclassical analysis in
Sec.~\ref{sec:semi}, and in Sec.~\ref{sec:results} we present our
results.  We close in Sec.~\ref{sec:conclude} with a summary of our
results and some concluding remarks.

\section{The Problem}
\label{sec:problem}

We study the motion of a particle of mass ${1/2}$ in a cosine potential
subject to a sinusoidally driven external force.  The Hamiltonian
is given by
\begin{equation}
H(p,x,t) = p^{2} - {1\over{2}} \cos{(\pi x)} 
                 - \epsilon x \sin{(\omega t + \phi )}
\label{Ham}
\end{equation}
with $\omega=2.5$, $\epsilon=0.126$ and $\phi=1.5493$.  We include the
nonzero constant $\phi$ \cite{foot1} to make contact with the
calculation of Ref.~\protect\onlinecite{me}.  For these choices of
$\omega$ and $\epsilon$ \cite{L&R} the external force may be considered
as a small perturbation in the context of KAM theory\cite{KAM}, and we
are justified in discussing the dynamics in the context of the remnant
orbit structure of the unperturbed motion.  The separatrix at the
threshold energy of the cosine potential plays a particularly important
role in the onset of the chaos, being the first orbit to rupture but
the last to fully disintegrate as a result of the perturbation.  This
orbit is shown in Fig.~\ref{fig_sep}(a) with the hyperbolic (unstable)
fixed points and the stable and unstable manifolds labeled
accordingly.

In this paper we study the dynamics for initial conditions in the
neighborhood of the point $(p_o,x_o)=(-0.7889,-0.3215)$.  The
trajectories starting in this neighborhood, being very near the
separatrix, rapidly lead to unbounded motion at $t_{ex} \approx 2.5$.
(Note that the exit time $t_{ex} \sim T$ where $T=2\pi/\omega$ is the
period of the external force.) This choice of initial conditions allows
us to examine the escape event associated with the onset of chaos with
a minimum of computational effort.

To study the quantum evolution, we propagate an initial wave function
forward in time by the split operator method\cite{Feit} generalized to
time-dependent systems.  Our choice of $\hbar=(200\pi)^{-1}$ sets the
quantum length scale for the problem to be small compared to the scale
of variation in the potential.  This allows us to pick an initial wave
function that is initially well-localized in both position and
momentum.  We choose for $\psi(x,0)$ a Gaussian wave packet of width
$\sigma$ centered at $(p_o,x_o)$
\begin{equation}
\psi(x,0) = (\pi\sigma^2)^{-1/4}
\exp{\left[-{(x-x_o)^{2}\over2\sigma^2}
 + i{p_o\over{\hbar}}(x-x_o)\right]}.
\label{GWP}
\end{equation}
The initial width $\sigma=0.0225$ of the wave packet is chosen to be
equal to that of the ground state wave function at the bottom of the
cosine potential.  We include one cosine well on each side of the well
at the origin (three wells total) to avoid spurious interference
between the escaping quantum amplitude and that remaining in the
initial well due to the periodic boundary conditions imposed by the FFT
algorithm.  A schematic depiction of the initial wave function in the
cosine potential is shown in Fig.~\ref{fig_sep}(b).  The solid curve
gives the cosine potential $V_o(x)$, while the dashed curve gives the
complete time-dependent potential $V(x,0)=V_o(x)-\epsilon
x\sin{(\phi)}$. Accompanying the seesaw motion of the external
washboard potential, the $x$ coordinate of each hyperbolic fixed point
at $(p_h,x_h)$ oscillates sinusoidally in time with $x_h \approx \pm
n-(2 \epsilon /\pi^2) \sin{(\omega t + \phi)}$ ($n=1,3,5...$) and
$p_h=0$.  The wave packet is launched in the negative $x$-direction, as
shown. After reflecting off the barrier at $x \approx -1$ (which rises
in time to meet it), it subsequently scatters off the top of the
barrier at $x \approx +1$ where it is partially transmitted.

To study the classical evolution, we evolve the classical equations of
motion forward in time for a set of 10~000 ``particles'' with different
initial conditions.  To ensure that the classical and quantum
descriptions agree initially, the initial conditions are drawn at
random from a two-dimensional Gaussian distribution of initial
conditions centered at $(p_o,x_o)$ and having widths in $p$ and $x$
equal to the widths in momentum and position of the initial wave
function of Eq.~(\ref{GWP}).  We compare the classical evolution to the
time-evolved quantum probability density $P_{qm}(x,t) =
|\psi(x,t)|^{2}$ by making a histogram with respect to $x$ of the
time-evolved classical distribution.

\section{Semiclassical Analysis}
\label{sec:semi}

We remind the reader that the time-evolved wave function is found from
the initial wave function $\psi(x,0)$ and the propagator as follows:
\begin{equation}
\psi(x,t) = \int \nolimits dx' \, G(x,x';t)\, \psi(x',0)
\label{int}
\end{equation}

For our semiclassical analysis, we require an accurate approximation
to the propagator.  Far from a caustic, we use the well-known
propagator due to Van Vleck and Gutzwiller\cite{vvg} 
\widetext
\begin{equation}
G_{VVG}(x,x',t) = \sum_{cl. paths}
\Biggl({1\over2\pi i\hbar}\Biggr)^{1/2}
\Biggl|{\partial^2 S(x,x',t)\over\partial x\partial x'}\Biggr|^{1/2}
                    \exp{\Biggl( i {S(x,x',t)\over{\hbar}}
                            -i\nu_{t}{\pi \over{2}}\Biggr)}\ ,
\label{G_VVG}
\end{equation}
\narrowtext
while close to a caustic we employ an expression for the propagator due
to Schulman\cite{Schulman} 
\widetext
\begin{equation}
G_{Sch}(x,x',t) = \Biggl({1\over2\pi i\hbar}\Biggr)^{1/2}
\Biggl| { {\em\lambda} (x,x',t)\over
(\partial^2 S/\partial x\partial x')^{-1}}\Biggr|^{1/2}_{x=x_c}
{\rm Ai}(z) \exp{\Biggl( i {S_c + p_c(x-x_c)\over \hbar}\Biggr)}.
\label{G_Sch}
\end{equation}
\narrowtext
In these two expressions $S(x,x',t)$ is the classical action as a
function of the initial ($x'$) and final ($x$) positions,
$S_c=S(x_c,x',t)$, and $(p_c,x_c)$ respectively give the action and
phase space coordinates at the caustic, and $\nu$ counts the number of
times the trajectory connecting $x'$ and $x$ encounters a caustic where
$\nu \pi /2$ is the ``Gutzwiller phase.'' The argument $z(x,x',t)$ of
the Airy function in Eq.~(\ref{G_Sch}) is found \cite{foot2} by solving
the boundary value problem for the quantum fluctuations around the
classical trajectory $x(x',t)$ \cite{Schulman} to obtain the
eigenvector with the lowest eigenvalue $\lambda_1 \propto {\em\lambda}
(x,x',t)/t^2$ While both the numerator and denominator in the prefactor
of Eq.~(\ref{G_Sch}) separately vanish at $x=x_c$, their ratio is
finite so that the singularity in Eq.~(\ref{G_VVG}) at $x=x_c$ is
absent.

Both of these expressions represent a WKB approximation to the exact
propagator\cite{Schulman}, meaning that the path integral is evaluated
by the method of steepest descents \cite{Chester}.  The former, which
assumes that the stationary paths are well separated, is valid when the
relative action for any two paths is much greater than $\hbar$, while
the latter, which assumes that two of the stationary paths are
coalescing at a caustic, is valid only in the immediate vicinity of the
caustic.

As an initial value problem, implementing the propagator of
Eq.~(\ref{G_VVG}) is a very reasonable task in one dimension for short
times, even when the dynamics are chaotic, and one can often get away
with simply ignoring the caustic spikes associated with the singularity
in the prefactor\cite{unexp}.  Unfortunately, as we mentioned above in
Sec.~\ref{sec:intro}, we do not find this to be the case for our
problem. But an exact calculation of Eq.~(\ref{G_Sch}) is basically
untenable because of the need to solve a boundary value problem at each
point in space.  To avoid this, we approximate $G_{Sch}$ by relating it
asymptotically to $G_{VVG}$.  The validity of this approximation rests
on being able to find a region not to close to the caustic where both
$G_{VVG}$ and $G_{Sch}$ are both reasonably accurate in order to match
them, similar in spirit to the usual WKB connection formulae for
semiclassical wave functions\cite{Schiff}.  This ``connection formula''
gives the argument of the Airy function in terms of the action
difference between the direct and reflected paths.  Thus, one has only
to solve the classical equations of motion (an initial value problem)
and compute the action along the classical paths to evaluate
$G_{Sch}$.

For our problem, the initial wave function of Eq.~(\ref{GWP}) is
sufficiently localized at $x'=x_o$ that near the caustic we may
approximate $\psi(x,t)$ in Eq.~(\ref{int}) by $G(x,x_o,t)$ directly.
In Equation (\ref{G_Sch}) we take $z$ to be sufficiently large and
negative (more precisely $|z|^{-3/2} \ll 1$) to replace ${\rm Ai}(z)$
with its leading asymptotic expansion.  Also, we assume that the sum
over classical paths in Eq.~(\ref{G_VVG}) consists of a single pair of
direct and reflected paths, and that the prefactors for these two paths
are equal.

In this approximation we find that
\widetext
\begin{equation}
|G_{VVG}(x,t)|^2 = 
{1\over2\pi\hbar}
\Biggl|{\partial^2 S(x,x',t)\over\partial x\partial x'}\Biggr|
_{x'=x_o}
\Biggl[1 + \cos{\left({\Delta S(x)\over\hbar} -{\pi\over 2}\right)}\Biggr]
\label{G2_VVG}
\end{equation}
\narrowtext
and
\widetext
\begin{equation}
|G_{Sch}(x,t)|^2 \propto
|z(x)|^{-1/2} 
\Biggl[1 + \cos\left({4\over3}|z(x)|^{3/2} - {\pi\over2}\right)\Biggr],
\label{G2_Sch}
\end{equation}
\narrowtext
where $\Delta S(x)=S_r(x,x_o,t)-S_d(x,x_o,t)\geq 0$. (The subscripts
$d$, $r$ denote the direct and reflected paths, respectively.) By
comparing the cosine arguments for Eqs.~(\ref{G2_VVG}) and
(\ref{G2_Sch}) one obtains the relation $z(x)=-(3\Delta
S(x)/4\hbar)^{2/3}$.  Although the condition $|z|^{3/2} \ll 1$ is
satisfied (to about a percent) only if $\Delta S/\hbar > 2\pi$, we find
that using this expression for $z$ in Eq.~(\ref{G_Sch}) works
remarkably well even as $\Delta S/\hbar \Rightarrow 0$. For the shadow
region of the caustic ($x>x_c$) we estimate the magnitude of $z=+|z|$
by reflecting the action difference about the caustic, {\em i.e.}, we
take $\Delta S(x)= \Delta S(x-x_c)$. We cannot infer the constant
prefactor of Eq.~(\ref{G_Sch}) in a similar manner because the quantity
$|z|^{-1/2} \approx (3\Delta /4\hbar)^{-1/3}$ only crudely mimics the
behavior of the VVG prefactor as $x \Rightarrow x_c$.

To evaluate $|G_{VVG}|^2$ and $|G_{Sch}|^2$ according to
Eqs.~(\ref{G2_VVG}) and (\ref{G2_Sch}) we evolve the vertical strip of
phase space defined by $x'=x_o$ forward in time according to the
classical equations of motion. At the desired time we obtain the action
$S(x,x',t)$ and its mixed partial derivative $\partial^2 S(x,x',t)
/\partial x\partial x' = -\partial p'(x,x',t)/\partial x$ as functions
of $x$ for the direct and reflected paths by interpolating between the
time-evolved grid points representing the strip.  Given these
quantities, the rest of the calculation is straightforward.

Although we have argued that the VVG propagator fails near caustics,
the overall character of the wave function is often preserved despite
the presence of caustic spikes. Even when the wave function is piled up
at the caustic, so that the effect cannot be ignored, the problem is
often temporary, disappearing once the amplitude has scattered away.
However, when there is a sustained accumulation of amplitude at the
caustic like we see in our problem, the VVG propagator becomes
inadequate.

To demonstrate this worst case scenario, we compare $|\psi(x,t)|^2$ for
the full quantum calculation to both the VVG and Schulman expressions
for $|G(x,x_o,t)|^2$ at $t=3$, as shown in Fig.~\ref{fig_demo}.  Both
curves are matched onto the full quantum calculation at the primary
maximum ($x \approx 0.89$) to determine the correct ``normalization''
constants.  This is necessary for the Schulman expression because we
don't know the constant prefactor. It is also necessary, in both cases,
because $\psi(x,0)$ isn't really a delta function.  For the sake of
comparison we also integrate Eq.~(\ref{int}) numerically for the VVG
propagator using Heller's Cellular Dynamics method \cite{cell}.  (We
cannot do the same for the Schulman propagator because our method of
approximation suppresses both the phase information and the prefactor.)
The critical and shadow regions of the caustic at $x_c \approx 1.09$
are indicated on the the plot.

Comparing the VVG and Schulman expressions for $|G(x,x_o,t)|^2$ to the
true quantum probability density, we see that the VVG expression gets
the oscillations to the left of the caustic about right, but it fares
badly in the critical region of the caustic, as expected, and it lacks
an exponential tail altogether. The Schulman expression, on the other
hand, not only gets the oscillations about right, but also correctly
describes the exponential behavior at the caustic and into the shadow
region where $z(x) > 0$. In the Cellular Dynamics calculation the
spurious singularity in the integrand causes the primary peak to have
too much amplitude and to be shifted to the right, and the change in
the location of the caustic across the set of time-evolved vertical
strips that contribute to the integral gives rise to a tail in the
shadow region that poorly approximates the true feature. While the
integration smears the singularity out, it clearly does not fix the
problem.

\section{Results}
\label{sec:results}

In order to discuss the quantum dynamics of this system in a
semiclassical context, it is necessary that we first point out several
features of the classical flow as it approaches and interacts with the
hyperbolic fixed points.  In Figures \ref{fig_dots}(a)-(h) we show the
time evolution of the classical phase space distribution for
$t=0.0-3.5$ at intervals of $\Delta t=0.5$ ($\approx 0.2T$).  Each dot
represents the time-evolved coordinates for one of the 10~000 particles
in the distribution.  We superimpose the separatrix of the unperturbed
problem in the first two frames for reference, and in the last two
frames we indicate both the caustic (at {\em c}) and the feature that
we call the false caustic (at {\em fc}).

Firstly, the initially compact object rapidly spreads out along the
remnant separatrix, being stretched by the unstable manifold of the
hyperbolic fixed point at $x \approx -1$, as shown in
Figs.~\ref{fig_dots}(a)-(d).  As a result, the particles appear to be
following the same (threshold) orbit in Fig.~\ref{fig_dots}(e). They,
in fact, approach the hyperbolic fixed point at $x \approx +1$ with a
fraction of the original energy dispersion. This behavior is distinctly
different from the spreading that occurs in the absence of the
time-dependent perturbation and is not just a consequence of the
orbital period being time-dependent.

Secondly, the bright side of the caustic ($x < x_c$) is not restricted
to the inside of the initial well ($x < x_h \approx 1$), as shown in
Figs.~\ref{fig_dots}(g) and \ref{fig_dots}(h).  The phase space
coordinates of the caustic are clearly outside the separatrix in both
cases.  The remnant separatrix is thus only a ``partial barrier to
transport'' \cite{MacKay} that temporarily confines the phase space
flow, and the interaction of the flow with the hyperbolic fixed point
is the door to the region of phase space that was inaccessible in the
absence of chaos. Here we note that the classical dynamics does not
achieve local mixing inside the remnant separatrix prior to the exit
event in contrast with the analysis of Bohigas {\em et al.}\cite{Bohig}
which assumes that local mixing is well established before regions
separated by partial transport barriers communicate with each other.

Thirdly, the tendril that results from the interaction of the flow with
the hyperbolic fixed point---meaning the feature in the Lagrangian
manifold between the caustic and the false caustic in
Figs.~\ref{fig_dots}(g) and \ref{fig_dots}(h)---gets thinner and longer
as a function of time. The tendril is thus acting like a closed curve
in phase space whose enclosed area is conserved by Liouville's theorem.
Neglecting the term $\int\nolimits \Delta H(t)dt \propto \epsilon$ that
follows from the inexact cancellation of the time-dependent potential
along the two paths\cite{me}, the area ``enclosed'' by the tendril
$\Delta A = \int_{x_c}^{x_{fc}} \Delta p(x)dx$ is indeed approximately
given by the relative action between the two paths at the false caustic
$\Delta S(x_{fc},x')$, where $(p_{fc},x_{fc})$ denote the coordinates
of the false caustic.

The false caustic is a fold in the Lagrangian manifold that does not
correspond to a turning point in the flow, meaning that the caustic
count $\nu$ is not incremented by one there, but rather, is decremented
by one. (In Figure \ref{fig_dots}(g) the fold is actually just on the
verge of forming.) Particles do not flow around this fold as a function
of time. Rather, those in the immediate vicinity of the false caustic
move rigidly with the manifold as it deforms. Consequently, the
quantity $\Delta S(x_{fc},x')$ approaches a constant value as the gap
between the two paths at the false caustic closes.  Since $\Delta
S(x_{fc},x')\approx \Delta A$, we see that the area-preserving nature
of the flow that gives rise to the stretching of the tendril is closely
tied to the existence of the false caustic in the Lagrangian manifold.

Having discussed some general features of the classical flow, we now
examine the corresponding quantum and classical probability densities
shown in Figs.~\ref{fig_prob}(a)-(h).  We see that the classical and
quantum distributions are essentially indistinguishable for $t < 2$ but
that this equivalence disappears once the tendril forms and the
classical motion becomes unbounded ($t > 2$). We should also point out
that, while some of the particles appear to escape the well to the left
at $t \approx 1$, they subsequently get pulled back inside (see
Figs.~\ref{fig_dots}(c)-(e) and Figs.~\ref{fig_prob}(c)-(e)).  This
spurious escape event, unlike the real event, is not accompanied by any
folding of the Lagrangian manifold, and the equivalence of the quantum
and classical distributions is preserved.

Just prior to the departure of the quantum from the classical behavior
at $t \sim 2.5$ we see that $x_c$ and $x_h$ are nearly coincident, as
are the quantum and classical maxima.  The caustic then moves off in
the positive $x$-direction away from the hyperbolic fixed point, and
the classical maximum follows.  The quantum maximum, however, is left
behind, stuck on top of the cosine barrier. This pinning of the quantum
peak on the barrier, as a mechanism for the inhibition of mixing, is
the key feature that we observe in the quantum chaotic dynamics of this
system.

We emphasize that this sticking or pinning phenomenon is not a result
of the difference in quantum and classical transmission coefficients
for a cosine barrier---which is negligible for our value of $\hbar$.
In particular, we find no appreciable difference between time-evolved
quantum and classical probability densities when we scatter a Gaussian
wave packet off an isolated, rigid cosine barrier at the threshold
energy.  Nor is the pinning caused by the side-to-side and/or
up-and-down motion of an isolated potential barrier.  If we mimic the
oscillations in position and energy near the hyperbolic fixed point of
the perturbed cosine potential with a potential barrier of the form
$(1+S(t))\cos{(kx-R(t))}$, we again find no appreciable difference
between the quantum and classical scattering at near-threshold
energies. We conclude that the history of the wavepacket in the cosine
well, particularly its delocalization prior to the escape event, is
crucial in giving rise to the differences between the classical and
quantum distributions that we observe.

Ironically, the delocalization of the wave packet is, in the following
sense, related to eigenfunction localization.  Having noted that the
classical phase space distribution of Fig.~\ref{fig_dots}(a) evolves to
resemble the separatrix, or at least part of it, one might expect the
corresponding time-evolved wave functions to resemble the
near-threshold eigenstates of the unperturbed Hamiltonian.  Quite
generally, for potentials with local maxima (or saddle points in two or
more dimensions) at energies well above the ground state energy, the
near-threshold eigenstates are highly peaked or ``localized'' at the
potential energy maxima\cite{Rau}. This feature is clearly manifested
in the wave functions of Figs.~\ref{fig_prob}(e)-(h).  The overlap of
the still localized wave packet and the extended threshold
eigenfunction is small in Figs.~\ref{fig_prob}(a) and
\ref{fig_prob}(b), but as the wave packet spreads out along the
separatrix, the overlap increases.  Thus, the accumulation of amplitude
at the hyperbolic fixed point represents the propensity for the chaotic
dynamics to increase the overlap between the evolving wave function and
the threshold eigenstates.  In this sense, the stickiness of the
hyperbolic fixed point that gives rise to the pinning effect is a
time-dependent manifestation of eigenfunction localization.

The pinning effect can also be understood dynamically as a quantum
interference effect between paths in the semiclassical propagator.  To
explain this we focus on the tendril feature in the region of the
caustic just after the escape event (at $t=3$).  We show the quantum
and classical probability densities as well as the Schulman/WKB
approximation of Eq.~(\ref{G2_Sch}) in Fig.~\ref{fig_detail}(a) and the
corresponding classical phase space distribution in
Fig.~\ref{fig_detail}(b).  In Figure \ref{fig_detail}(c) we plot the
relative action $\Delta S(x)$ and VVG path amplitude $|\partial^2 S/
\partial x \partial x'|^{-1/2}_{x'=x_o}$ for the time-evolved vertical
strip $p(x,x_o)$.  The direct and reflected paths and the locations of
the caustic ({\em c}), false caustic ({\em fc}) and $x$-coordinate of
the hyperbolic fixed point are indicated accordingly.

The agreement between the quantum distribution and the WKB
approximation in Fig.~\ref{fig_detail}(a) clearly demonstrates that the
structure of the wave function in the vicinity of the tendril has a
semiclassical origin.  Specifically, the oscillatory nature of the wave
function here is caused by interference between the two nearby paths
that comprise the tendril.  These paths do not contribute to the
propagator independently, however, because of the proximity of their
actions (see Fig.~\ref{fig_detail}(c)), which is why we must use
Schulman's approximation to the two-path WKB propagator to describe the
interference properly.  Our semiclassical analysis is only valid in the
region $x > x_{fc}$, however, because of the sudden divergence of the
VVG amplitudes to the left of the false caustic (see also
Fig.~\ref{fig_detail}(c)).  We observe similar agreement between the
quantum distribution and Schulman's approximation at $t=2.5$ and
$t=3.5$, although the presence of additional paths complicate matters
somewhat in the latter case.

The region for which $ \Delta S(x,x')/\hbar < 1$---or equivalently, the
region where Schulman's propagator should replace the VVG
propagator---grows directly with the length of the tendril.  Since the
area enclosed by the tendril $\Delta A$ is roughly constant in time
with $\Delta A \approx \Delta S(x_{fc},x')$, as explained above, it
follows that the Airy structure associated with the tendril simply gets
stretched as the tendril is stretched.  As a result, the quantum peak
remains near $x_h \approx +1$ where it originated.  The area-preserving
nature of the chaotic Hamiltonian flow, coupled with the presence of
the false caustic in the Lagrangian manifold, is thus responsible for
the peak in the quantum distribution being held at the hyperbolic fixed
point, thereby inhibiting the quantum transport of probability across
the separatrix.

\section{Conclusions}
\label{sec:conclude}

We have shown that the quantum probability density rapidly diverges
from that of the analogous classical system when the tendril formed by
the interaction of the Lagrangian manifold with the hyperbolic fixed
point leads to (classically) unbounded motion.  We have also shown that
this divergence is the consequence of a quantum interference effect
that pins the quantum peak associated with the caustic at the top of
the potential barrier.  This pinning effect is a concrete example of
how quantum interference suppresses the exploration of ``phase space''
outside the broken separatrix to inhibit the classical mixing.

The origin of the pinning effect was shown to be semiclassical,
resulting from interference in the propagator between the direct and
reflected paths associated with the tendril. We found it necessary to
use an approximation to the WKB propagator due to Schulman that is
valid near caustics, rather than the Van Vleck-Gutzwiller propagator,
in order to study this interference properly. This is because the
stretching of the tendril that accompanies the chaotic dynamics phase
locks the entire region of phase space explored by the tendril in close
proximity to the caustic. We approximate Schulman's propagator by
relating its asymptotic expansion to the Van Vleck-Gutzwiller
propagator.  This avoids the untenable boundary value problem one must
otherwise solve in order to calculate Schulman's propagator directly.
We have shown that Schulman's propagator, evaluated in this way,
accounts for the difference between the behavior of the quantum and
classical systems at the times we consider.

We emphasize that this interference occurs between direct and reflected
paths, differing in Gutzwiller phase by $\pi/2$---in contrast with the
interference effect discussed in Ref.~\onlinecite{me}.  Thus, we
conclude that the interference between direct and reflected paths (for
these very early times) has no permanent effect on the wave function as
it sloshes back and forth in the potential well as long as the
classical motion remains bounded---just as the structure that appears
in a wave packet as it scatters off of a rigid barrier\cite{Schiff} is
temporary, disappearing once the packet is fully reflected.  When the
motion becomes unbounded, however, we have shown that the interference
between direct and reflected paths in the tendril permanently modifies
the wave function, profoundly affecting the quantum-classical
equivalence that characterizes the dynamics prior to the exit time.  We
have shown that this divergence of the quantum and classical behavior
is a direct consequence of the folding and stretching of the Lagrangian
manifold that accompanies the classical flow across the separatrix; and
this, in turn, is a direct consequence of the onset of the classical
chaos.

Because the pinning effect is directly related, in the semiclassical
sense, to generic features of the classical Hamiltonian chaos, we argue
that this mechanism for the inhibition of mixing is itself generic.
Thus, one should expect to find the pinning effect in other weakly
driven chaotic Hamiltonian systems as well.  Moreover, since classical
canonical perturbation theory and the KAM theory for the origin of soft
chaos are applicable for both time-dependent and time-independent
perturbations\cite{Chirikov}, this mechanism should also appear in
conservative (Hamiltonian) systems characterized by soft chaos.
Lacking the time-dependence responsible for the false caustic in our
problem, we suspect that the higher order caustics (where the caustic
count $\nu$ changes by more than one) that can exist in such systems
because of the increased dimensionality would play the role of the
false caustic in giving rise to the pinning effect.

We cannot go so far as to argue that the pinning effect is generic to
all Hamiltonian quantum chaos, however, because the presence of
tendrils and either false caustics or (possibly) higher order caustics
in the Lagrangian manifold is not, by itself, sufficient to produce
this feature. The partial barrier to classical phase space transport
associated with the broken separatrix appears to be the more crucial
element. While the strongly chaotic stadium billiard problem, for
example, exhibits both higher order caustics \cite{areahbar} and
tendril-like structures (see the two-dimensional Birkhoff projection of
the Lagrangian manifold in Ref.~\onlinecite{areahbar}), it does not
possess the simple KAM structures ({\em i.e.}, the cantori or broken
separatrices) that act as partial barriers to transport\cite{MacKay}.
Nor does the quantum system manifest the stretching of the exponential
tail in the classically forbidden region (or the associated pinning
effect), or so we assume given the long-time accuracy of the VVG
propagator achieved by Heller and Tomsovic for that problem mentioned
in Sec.~\ref{sec:intro}.  But whether or not this mechanism is
significant in other systems characterized by hard chaos is still an
open question.

\acknowledgments

We would like to thank Steve Tomsovic and Harold Baranger for useful
discussions.  We would also like to thank Vijay Poduri for bringing our
attention to Ref.~\onlinecite{Schulman}.  This work was supported by
the National Science Foundation under Grant No.\ DMR--9020310.

\begin{figure}
\caption{
(a) The unperturbed separatrix at zero energy showing the stable and
unstable manifolds of the hyperbolic fixed points at $x=\pm 1$.
(b) The unperturbed (solid) and perturbed (dashed) potentials
$V_o(x)=\cos{(\pi x)}$ and $V(x,0)=V_o(x) - \epsilon x \sin{(\phi)}$
with a schematic depiction of the initial wave packet centered at
$(p_o,x_o)=(-0.7889,-0.3215)$.
}
\label{fig_sep}
\end{figure}

\begin{figure}
\caption{
A comparison of $P_{qm}(x,t)$ (solid curve) to $|G(x,x_o,t)|_{VVG}^2$
(dash-dot curve) and $|G(x,x_o,t)|_{Sch}^2$ (dashed curve) at $t=3$.
Also shown is the Cellular Dynamics calculation of the VVG probability
density (dotted curve).
}
\label{fig_demo}
\end{figure}

\begin{figure}
\caption{
(a)-(h) The time-evolution of the Gaussian phase space distribution
at increments of $\Delta t=0.5$.
The ``tendril'' as discussed in the text refers to the feature
in the distribution between the caustic (at {\em c}) and the
false caustic (at {\em fc}) in (g) and (h).
}
\label{fig_dots}
\end{figure}

\begin{figure}
\caption{
(a)-(h) The time-evolution of $P_{cl}(x,t)$ (dashed curve) and
$P_{qm}(x,t)$ (solid curve) at increments of $\Delta t=0.5 \approx
0.2T$.  Note the divergence of the classical and quantum maxima that
occurs in (g) and (h) accompanying the emergence of the tendril in
Fig.~\protect\ref{fig_dots}.
}
\label{fig_prob}
\end{figure}

\begin{figure}
\caption{
(a) The same curves of Fig.~\protect\ref{fig_prob}(g) magnified to show
the structure associated with the tendril. Also shown is the
semiclassical approximation $|\psi|^2 \sim |G_{Sch}|^2$ as described in
Sec.~\protect\ref{sec:semi} (dotted curve).
(b) The corresponding phase space distribution of
Fig.~\protect\ref{fig_dots}(g).
(c) The relative action $\Delta S(x)/2 \pi$ in units of $\hbar$ (solid
curve), and the (inverse) VVG amplitude $\delta S(x) \equiv
\protect\sqrt{2\pi\hbar}\, |\partial^2 S/\partial x \partial
x'|^{-1/2}_{x'=x_o}$ along the direct and reflected paths (dashed
curves).
}
\label{fig_detail}
\end{figure}


\begin{references}
\bibitem{foot1}
In order to amplify the interesting behavior associated with the
classical exit event for the initial conditions of
Ref.~\protect\onlinecite{me}, we followed the unstable classical
trajectory used in Ref.~\protect\onlinecite{me} up to a time $t=12/\pi$
shortly before the classical exit event and launched a localized wave
packet there.
\bibitem{foot2}
In Schulman's notation the prefactor in Eq.~(\protect\ref{G_Sch}) is
given (up to an overall time-dependent ``constant'') by
$\sqrt{\lambda_1 /f(t)}$.  $\lambda_1$ is found by solving the equation
for the Jacobi field $\phi$ $m\ddot{\phi} + \partial^2 V/\partial
x^2|_{x_{cl}} \phi + \lambda_n \phi = 0$ with $\phi(0)=\phi(t)=0$ and
$\lambda_1 < \lambda_2 < ...$, while $f(t)$ is found from the initial
value problem $m\ddot{f} + \partial^2 V/\partial x^2|_{x_{cl}} f = 0$
with $f(0)=0$ and $\dot{f}(0)=1$. Our form for the propagator in
Eq.~(\protect\ref{G_Sch}) follows given that $f(t) \equiv -m(\partial^2
S/\partial x \partial x')^{-1}$.  (See
Ref.~\protect\onlinecite{Schulman} ch. 12-13,~15.)
\end{references}
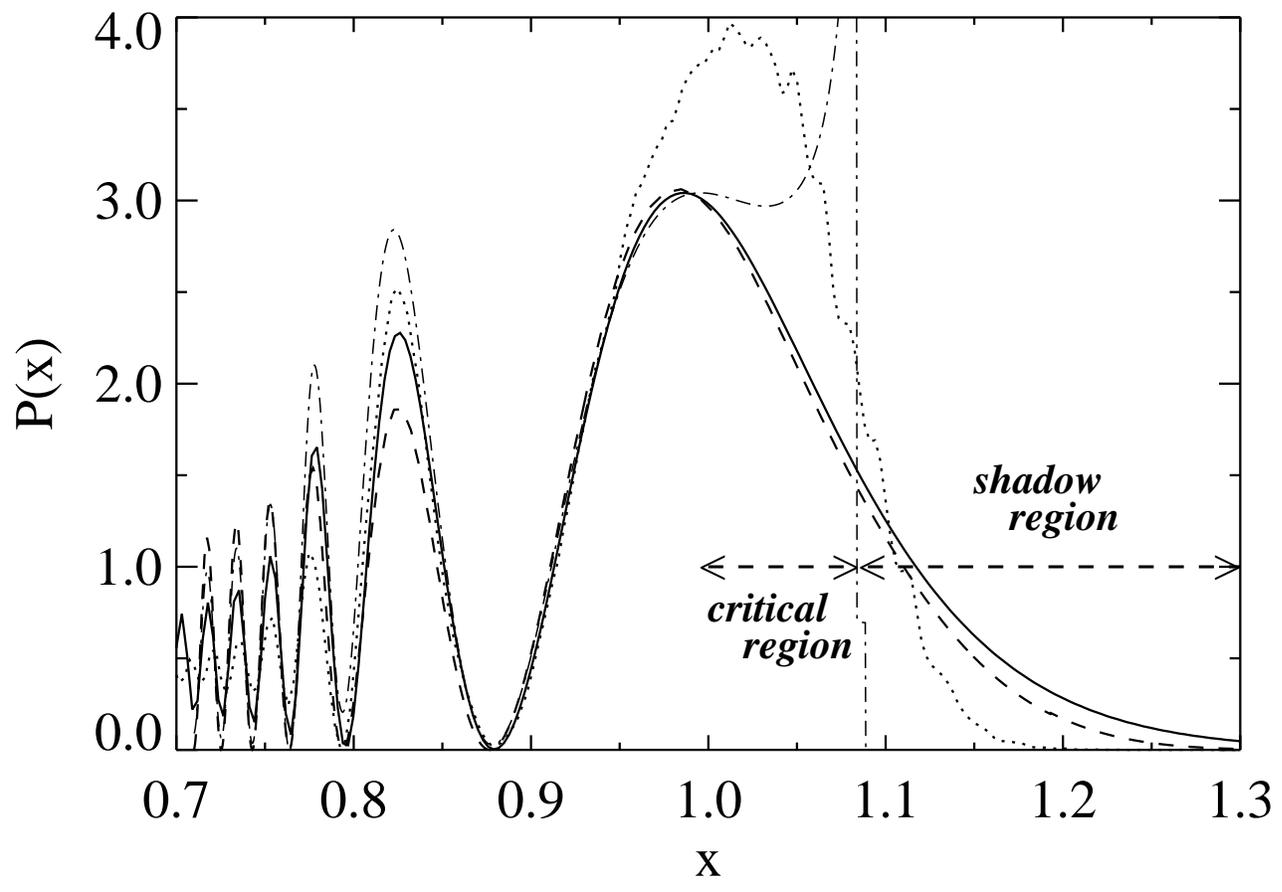
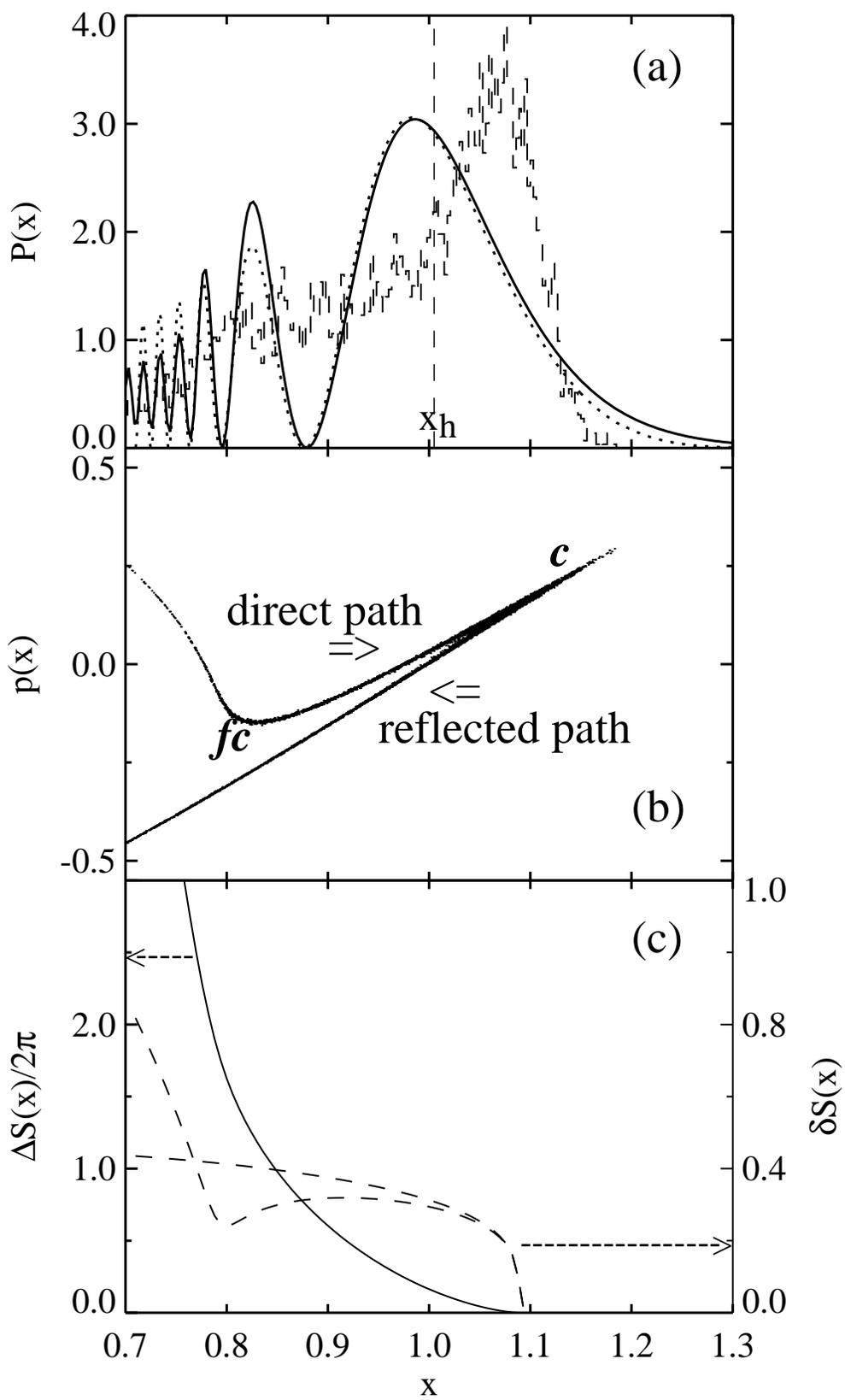
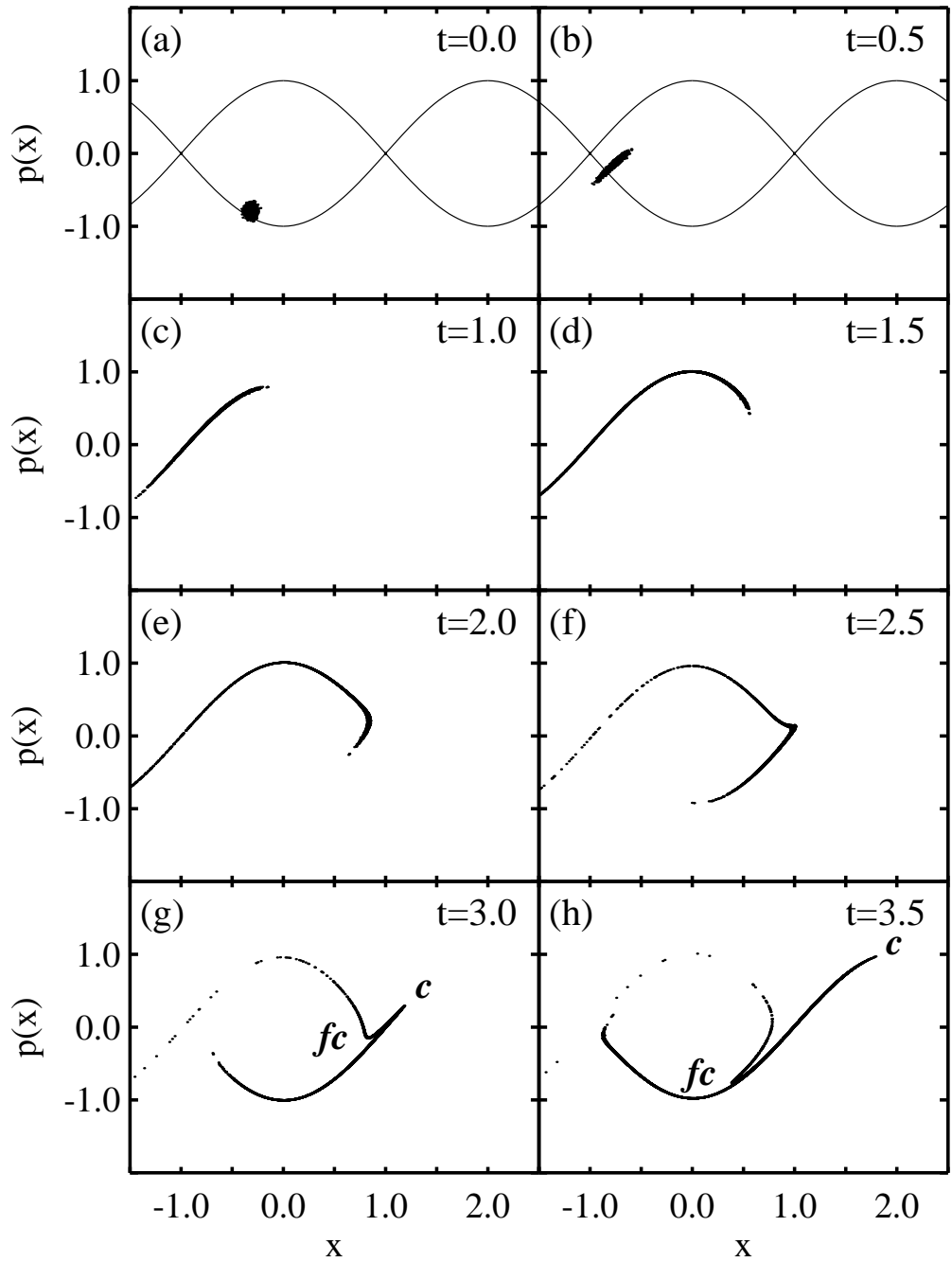
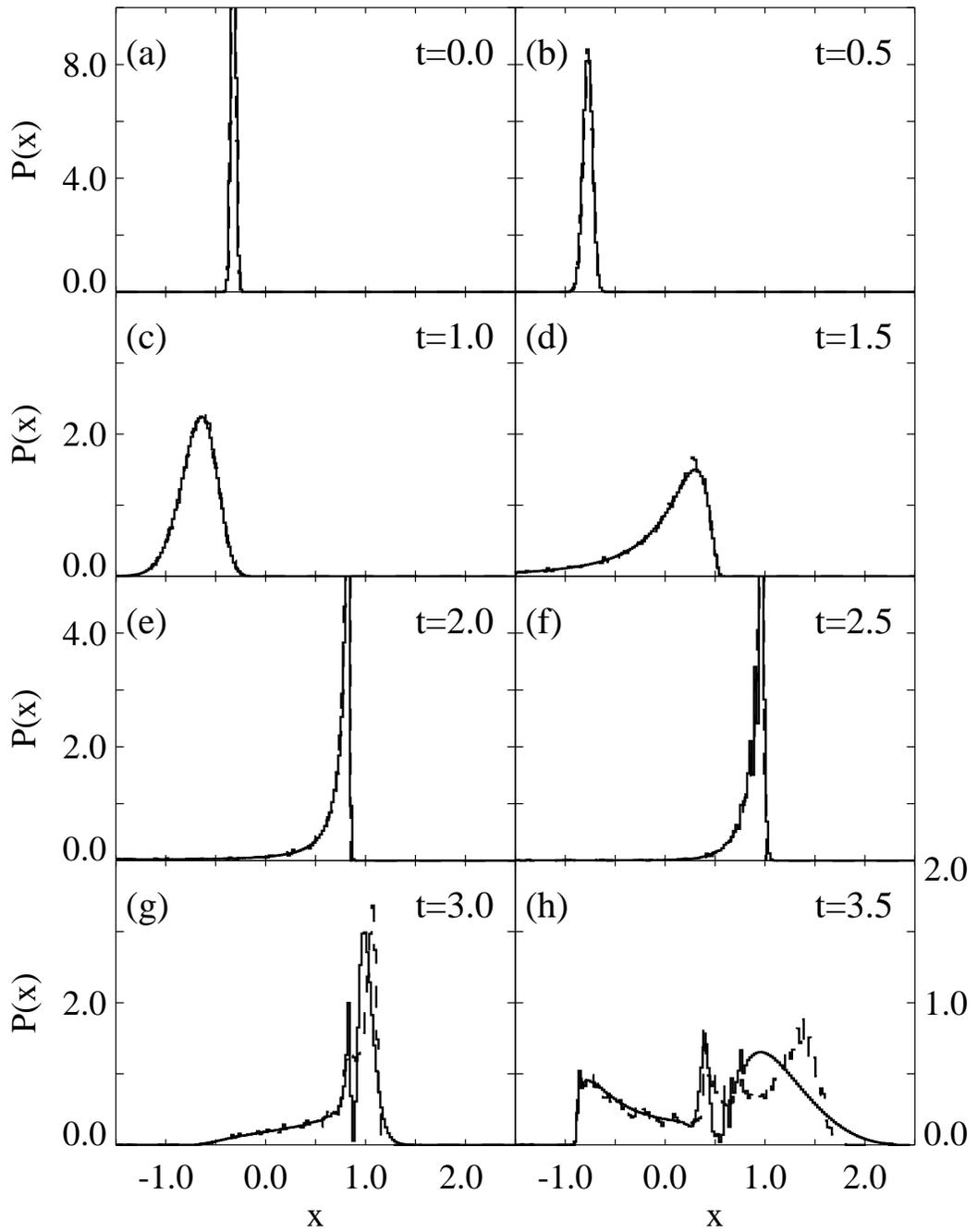
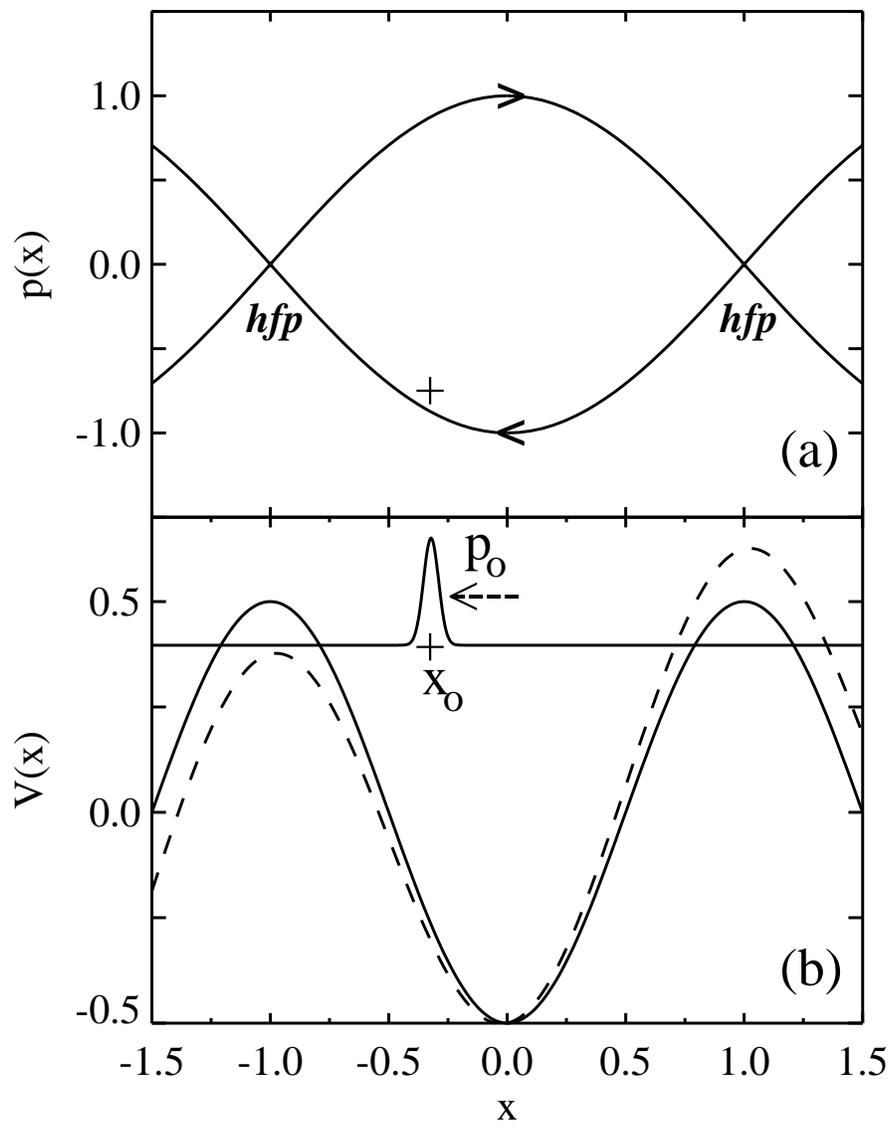
\end{document}